\documentclass[prl,floatfix,twocolumn,showpacs,amsmath,amssymb]{revtex4}
\usepackage{graphicx,dcolumn,bm,color}

\begin{document}

\title{Magnetic field-induced transition in a quantum magnet described by the Quantum Dimer Model}
\author{Arnaud Ralko,$^{1}$ Federico Becca,$^{2,3}$ and Didier Poilblanc$^{1}$}
\affiliation{
$^{1}$ Laboratoire de Physique Th\'eorique, Universit\'e Paul Sabatier, CNRS, 
F-31400 Toulouse, France \\
$^{2}$ International School for Advanced Studies (SISSA),
Via Beirut 2, I-34014 Trieste, Italy \\
$^{3}$ CNR-INFM-Democritos National Simulation Centre, I-34014 Trieste, Italy.
}

\date{\today}

\begin{abstract}
The effect of a magnetic field on a gapped quantum magnet is described within 
the framework of the Quantum Dimer Model. A minimal model describing the 
proliferation of itinerant spinons above a critical field is proposed and 
investigated by Lanczos exact diagonalizations and quantum Monte Carlo 
simulations. For both square and triangular lattices, it is shown that 
spinons are fully polarized and Bose-condense. This offers a novel scenario 
of a Quantum Critical Point in the dimer-liquid phase (triangular lattice) 
characterized by the continuous appearance of a spinon superfluid density, 
contrasting with the usual triplet condensation picture. The possible role of 
other spinon kinetic terms neglected in the model are discussed.
\end{abstract}


\maketitle

Quantum (spin) disordered magnets are currently a very active field of
research in Condensed Matter Physics. They are characterized by the absence of
magnetic ordering down to very low temperatures and, for some of them of
interest here, by a spin gap to triplet excitations. In this context, the
magnetic field is a particularly relevant external parameter as it can drive
the system to a Quantum Critical Point (QCP) at which the spin gap
vanishes~\cite{giamarchi}.
In the conventional picture, the QCP can be viewed as the Bose condensation of
triplets. Numerous examples of such QCP can be found in quasi-one dimensional
magnets like the CuHpCl spin ladder ~\cite{chaboussant}, in two dimensional
layered frustrated magnets like e.g. the Copper-Borate 
system~\cite{CuBO3} or in the interesting three dimensional BaCuSiO
dimer compound~\cite{sebastian}.

The resonating valence bond (RVB) liquid describes the paradigm for 
magnetically disordered systems. Indeed, RVB states describe pure quantum 
phases, which have no classical counterparts and, therefore, cannot be 
adiabatically connected to any band theory, based on Hartree-Fock or more 
sophisticated density functional approaches. Moreover, as suggested by 
Anderson in a cornerstone paper~\cite{anderson}, an RVB insulator could be 
intimately connected to high-temperature superconductors. In this field,
an important step forward was done by Rokhsar and Kivelson, who formulated 
the RVB concept by introducing a simple microscopic Hamiltonian, the so-called 
Quantum Dimer Model (QDM)~\cite{rokhsar}.
Simplifying the original electronic problem, they proposed an effective theory 
describing the dynamics of short-range dimers, that become the fundamental 
entities of the problem. The QDM relies on two approximations which 
are believed not to alter significantly its physical relevance (and could be 
relieved at the price of bringing an enormous extra complexity):
(i) only nearest-neighbor singlets are considered and (ii) different dimer 
configurations are taken to be orthogonal to each other, hence reducing 
singlets to ``classical'' dimers. However, quantum mechanics still enters in 
an essential way in the possibility to flip plaquettes with parallel dimers.
On the square lattice, the ground state of the QDM always shows lattice
symmetry breaking. This fact has encouraged the (wrong) surmise that ``fully''
disordered ground state (in the sense of a state with neither SU(2) nor 
lattice symmetry breaking) cannot be stabilized in this framework. 
In this respect, the discovery that a gapped and disordered state can be 
obtained on the triangular lattice opened the way to an enormous number of
investigations~\cite{moessner,moessner3}. Recently, the zero-temperature
properties of the QDM on the triangular and square lattices have been
considered in great detail by numerical calculations~\cite{ralko1,ralko5}.

In the QDM, both charge and spin excitations are frozen and the elementary 
excitations have topological character~\cite{read,ivanov,ralko3}.
Some effort has been done in order to introduce charged spinless excitations
in the doped QDM, the so-called holons. The exact nature of these objects has
been widely debated~\cite{read,kivelson,poilblanc}. 
Much more subtle is the problem of inserting spinons, namely chargeless
spin-1/2 objects, since, in this case, one faces a serious problem of
non-orthogonality~\cite{read}. In fact SU(2) valence-bond configurations with 
unpaired spins at different locations are not orthogonal to one another, in 
contrast to the holon case. The magnetic field offers a formidable tool to 
experimentally investigate such issues since it plays the role of a chemical 
potential w.r.t. the spinon density. 

In this Letter, we address the issue of spinon doping in both (lattice) 
ordered and disordered dimer backgrounds. Physically, this is achieved by the 
action of a magnetic field. On a general basis, we first derive all types of 
terms that should govern the spinon dynamics. Then, we propose a simplified 
model where the (up or down) spinon can hop along plaquette diagonals 
(while its neighboring dimer rotates by 90 degrees) and where the dimer 
dynamics is that of the standard QDM. 
By using Lanczos exact diagonalizations, we show that, within this model, 
for a finite magnetic field, all spinons are fully polarized, the energy gain 
being of kinetic origin as in the well known Nagaoka effect~\cite{nagaoka}.
The magnetization profiles are then computed using Green's function Monte 
Carlo. We find a new type of QCP characterized by Bose-Condensation of
spinons in contrast to the usual scenario of triplet condensate. The
behavior of the magnetization at the QCP depends on the nature of the (spin)
disordered ground state at zero field.

\begin{figure}
\includegraphics[width=0.92\columnwidth,clip]{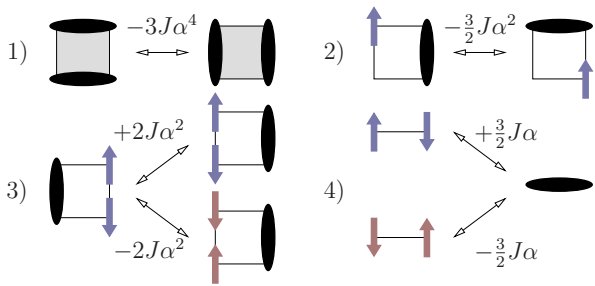}
\caption{\label{fig1}  
(Color online) The four different kinetic processes of the effective 
Hamiltonian and their corresponding matrix elements in the case of a 
(non-frustrated) Heisenberg quantum magnet. $\alpha = 1/\sqrt{2}$ is
the singlet normalization factor and $J$ the spin coupling. The sign
convention for dimers is the one of Ref.~\onlinecite{rokhsar}.}
\end{figure}

Creating a pair of spinons in a gapped quantum magnet (described by a QDM) 
will cost a finite energy (called $\Delta_B$ later on). However, a finite 
density $x$ of (partially polarized) spinons could be stabilized by the Zeeman 
energy $- g \mu_B H \sum_i S_i^z = 
-\frac{h}{2}N \left( n_{\uparrow} - n_{\downarrow}\right)$,
$n_{\uparrow}$ ($n_{\downarrow}$) being the densities of up (down) spinons in
the system, $x = n_{\uparrow} + n_{\downarrow}$ and $N$ is the number of sites. 
Therefore, as a preliminary work, we need to establish the form of the kinetic 
processes for the spinons moving in the fluctuating dimer background.
At the microscopic level, such terms will {\it a priori} arise from the 
non-orthogonality of the basis set. Since the derivation for a true frustrated 
microscopic model is beyond the scope of this work, we simply consider here 
the (non-frustrated) Heisenberg model on the square lattice to get insights 
on the {\it generic} form of these matrix elements~\cite{explain}.
Notice that a similar type of derivation has been curried out for a realistic 
but more complex model in a different context~\cite{vernay}.

Following the same procedure as for the derivation of the QDM in the absence 
of spinons~\cite{kivelson}, we calculate the largest contributions of the 
matrix elements, $H_{c,c'} = \langle c' | H_{\textrm{Heis}}| c \rangle$, 
where $|c \rangle$ and $|c' \rangle$ are typical dimer configurations, 
including spinons. A careful analysis gives four elementary kinetic processes 
and their overlap weights, as displayed in Fig.~\ref{fig1}. Interestingly, 
the processes involving a spin-flip of two spinons [term 3)] or two-spinon 
creation or annihilation [term 4)] have both positive and negative signs 
depending on the orientation of the spinons. This introduces a serious sign 
problem into the quantum Monte Carlo approach. In addition, the term 4), 
in contrast to the holon-doped case, does not conserve the number of 
``dopants''.

Hereafter, we consider a simplified version of the QDM-spinon model, by 
retaining only the dimer flips (as in the usual QDM) and the spinon
hopping 2). This model has only negative off-diagonal matrix elements and 
should be tractable by appropriate quantum Monte Carlo. In a second step, 
we shall discuss the possible role of the terms we have left behind. 
Note that in the case of a frustrated magnet, we expect the same type of 
processes although their magnitudes should depend on the precise microscopic 
magnetic interactions.  Therefore, the QDM Hamiltonian is defined here by
\begin{equation}\label{hamilt}
{\cal H}_{0} = v \sum_{c} N_c |c \rangle \langle c| - 
J \sum_{(c,c')} |c' \rangle \langle c| 
-t \sum_{(c,c'')} |c'' \rangle \langle c|.
\end{equation}
The sum on $(c)$ runs over all configurations of the Hilbert space, 
$N_c$ is the number of flippable plaquettes, the sum on $(c',c)$ runs over 
all configurations $|c \rangle$ and $|c' \rangle$ that differ by a single 
plaquette dimer flip, and the sum on $(c'',c)$ runs over all configurations 
$|c \rangle$ and $|c''\rangle$ that differ by a single spinon hopping between 
nearest neighbors (triangular lattice) or next-nearest neighbors 
(square lattice), see Fig.~\ref{fig1}. The coupling to the external magnetic 
field is also added to the above Hamiltonian.

\begin{figure}
\includegraphics[width=0.7\columnwidth,clip]{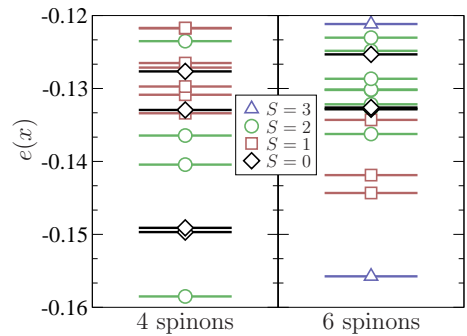}
\caption{\label{fig2}
(Color online) The lowest-energy levels for four and six spinons on a 
$4 \times 4$ square lattice in the subspace of zero momentum. The couplings are
$t/J=0.2$ and $v/J=0.5$ and states are classified according to the total spin
$S$.}
\end{figure}

To get insights on the ground-state properties of this system, we first
exactly diagonalized (by standard routines or a Lanczos algorithm) a 
$4\times 4$ square cluster with periodic boundary conditions and with 4 and 6 
doped spinons. Eigen-energies are labeled according to the total spin quantum 
number~\cite{nota2spin}. Results are reported in Fig.~\ref{fig2} for zero 
magnetic field ($h=0$). The important point is that, in all the cases we 
considered, the ground state has the maximum spin value, namely it is fully 
polarized. This bears strong similarities with conventional Nagaoka 
ferromagnetism~\cite{nagaoka} (but for a finite concentration of dopants, 
which is remarkable) in the sense that the energy gain w.r.t. the singlet 
subspace is of kinetic origin. Obviously, the fully polarized state is 
further stabilized by a finite magnetic field so that this property should be 
valid for all fields and dopant concentration making quantum Monte Carlo 
simulations available. Note however that the neglected processes 3) and 4) of 
Fig.~\ref{fig1} might change the relative stability of the fully polarized 
state w.r.t. some lower spin states and this will be discussed later on. In 
this way, technically, the spin variable does not play any role and, therefore,
one can borrow here the results obtained in the context of the monomer-doped 
QDM~\cite{poilblanc2,ralko4}, but accounting for the extra term involving the 
magnetic field $-hN x/2$ ($n_{\uparrow}=x$). It is of interest to notice that, 
in contrast to the case of hole-doping where Fermi statistics plays 
a role~\cite{poilblanc}, ``spinon-doping'' do not introduce any minus-sign
problem (here $J>0$). In other words, since an electron destruction operator 
can be written as $c_{i\sigma}=f_i^\dagger b_{i\sigma}$ in term of a fermionic 
holon-creation operator $f_i^\dagger$ and a bosonic spinon-destruction operator
$b_{i\sigma}$, the physical holon-doped and ``spinon-doped'' QDM simply differ 
from the statistic of the {\it bare} doped monomers.

Here, we aim to assess the number of spinons that are present in the 
ground state, once an external magnetic field $h$ is applied and competes with
the dimer binding energy $\Delta_B$. Indeed, $-h$ plays the role of a chemical 
potential and controls the spinon density which gives the reduced 
magnetization. It is natural to introduce the effective energy cost
$\Delta_{B}^{\textrm{eff}} = \Delta_B - h$ of breaking a dimer into two unbound 
{\it static} spinons under a finite magnetic field. Note that the spontaneous 
appearance of spinons in the ground-state (which defines the QCP) should 
appear for fields {\it lower} than the characteristic field  $h=\Delta_B$ 
since spinons acquire extra kinetic energy compared to dimers (in other words, 
the actual critical value of $\Delta_{B}^{\textrm{eff}}$ is positive).
This study can be done by considering the energy per site $e(x)$ of the 
holon-doped model~\cite{ralko4}, from which the Legendre transform 
$F(x) = e(x) +\frac{1}{2}\Delta_{B}^{\textrm{eff}}x$ can be constructed.
The actual $e(x)$ has been obtained by using the Green's function Monte Carlo 
method~\cite{nandini}, for both the square and the triangular 
lattices~\cite{ralko4}.
The actual spinon concentration $x_0$ for a given value of
$\Delta_{B}^{\textrm{eff}}$ is obtained by minimizing $F(x)$ with respect to
$x$. Notice that $x_0$ is nothing else but the ratio of the magnetizations
$M/M_{\rm{sat}}$, with $M_{\rm{sat}}$ the value at saturation, i.e., when the
whole system is filled by spinons. 

\begin{figure}
\includegraphics[width=0.9\columnwidth,clip]{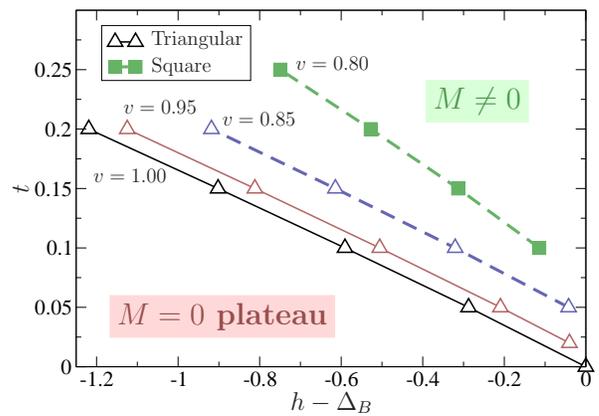}
\caption{\label{fig3}
(Color online) Phase diagram of the QDM  given by Eq.~(\ref{hamilt}) as a 
function of magnetic field, for the square lattice (full squares) and the 
triangular lattice (empty triangles). The dashed and continuous lines indicate 
first- and second-order transitions, respectively.}
\end{figure} 

The fact that the monomer-doped model may or may not show phase separation
(PS), opens the way to two different scenarios for the behavior of the system
as a function of the magnetic field. Let us first recall that, on the square 
lattice, as soon as $v/J<1$, the ground state is unstable in the presence 
of an infinitesimal monomer concentration and phase-separates between undoped 
(macroscopic) domains and regions with a concentration $x_c$ of monomers. 
Instead, in the triangular lattice, the infinitesimally doped system is stable 
for $v/J \gtrsim 0.9$, whereas it phase-separates for smaller values of the 
dimer-dimer repulsion~\cite{ralko4}. Such different behaviors are reflected 
in the phase diagram of Fig.~\ref{fig3} in the $(h-\Delta_B,t)$ plane by continuous 
or first-order transitions between the regions with and without magnetization. 
Remarkably, the transition curves $t=t^c(h)$ have linear behaviors and, for a 
given lattice structure, the slope does not depend upon the ratio $v/J$. 
Three different values of the dimer-dimer repulsion $v/J$ have been chosen 
for the triangular lattice, two of them inside the stability region of the 
RVB liquid of the undoped system ($v/J=1$ and $0.95$), whereas the latter 
one ($v/J=0.85$) corresponds to a region where the doped ground state phase 
separates, due to local dimer ordering.

\begin{figure}
\includegraphics[width=0.9\columnwidth,clip]{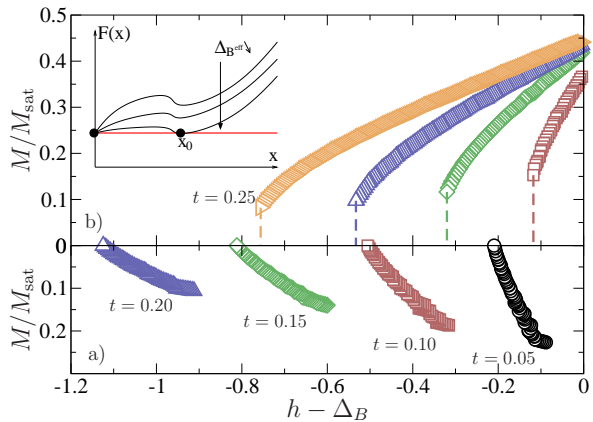}
\caption{\label{fig4}
(Color online) a) Spinon concentration for different values of $t$ as a
function of the magnetic field $h$ (both in units of $J$) for the triangular
lattice at $v=0.95$. b) The same as in a), but for the square lattice and at
$v=0.80$. Dashed lines show the region where a first order transition occurs.
Error-bars are of the size of the symbols.}
\end{figure}

The behavior of the magnetization vs magnetic field can be understood on 
simple grounds. In a stable system, the compressibility is finite, implying 
that the second derivative of $e(x)$ is positive (i.e., $e(x)$ is a concave up 
function of $x$). In this case, $e(x) \simeq e(0)+ax+bx^2/2$, where $b>0$ 
from the stability criterion. Moreover, the linear coefficient is generally 
negative, i.e., $a=-|a|$. On the other hand, PS is signaled by an infinite 
compressibility and a vanishing second derivative of $e(x)$. This leads to a 
precise form of the energy in the thermodynamic limit, namely $e(x)=e(0)+ax$ 
for $x<x_c$ and $e(x) \simeq e(x_c) + a(x-x_c) + b(x-x_c)^2/2$ for $x \ge x_c$.
Notice that the correct behavior of the stable case may be found from the
previous one by imposing $x_c=0$. By minimizing $F(x)$ we then find a general 
expression for the reduced magnetization,
\begin{equation}
M/M_{\rm sat} = x_c +\frac{|a|}{b} + \frac{1}{2b}(h-\Delta_B),
\end{equation}
for $h > 2|a| - \Delta_B$ and $M=0$ otherwise. Therefore, in a stable system
(i.e., $x_c=0$), by increasing the magnetic field, there is a continuous 
transition between the gapped phase to a phase with a finite spinon density. 
In Fig.~\ref{fig4}a, we show the behavior of $x_0=M/M_{\rm sat}$ as a function of $h-\Delta_B$ 
for the triangular lattice and $v/J=0.95$, namely starting from a $h=0$ RVB 
liquid phase.
A Bose-condensation of spinons is found, characterized by a second-order 
phase transition for all the values of $t/J$ considered, hence confirming
the analytical arguments.     

On the contrary, whenever $x_c \ne 0$, by increasing the magnetic field, the 
number of spinons suddenly jumps to a finite value for $h>\Delta_B-2|a|$. 
In this case, the critical value of the corresponding (effective) dimer 
binding energy is given by $\Delta_{B,c}^{\textrm{eff}}=2|a|$, i.e., twice the 
linear slope of the energy per site at zero doping, $|a|=[e(x_c)-e(0)]/x_c$.
In panel Fig.~\ref{fig4}b, we consider the case of the square lattice, with
$v/J=0.8$. At small enough magnetic field, the ground state is an anisotropic 
mixed columnar/plaquette phase, as recently pointed out~\cite{ralko5}. 
Upon increasing $h$, the ground state exhibits a first-order transition at 
which $M/M_{\rm{sat}}$ jumps suddenly to a finite value. 

Though serious ergodicity problems prevent us to consider very small values
of $t/J$, we have evidence that, except for $v/J=1$, the transition lines in
Fig.~\ref{fig3} extrapolate to non-zero values of $t/J$ as $h \to \Delta_B$.
This fact implies that the linear coefficient of the energy per site $e(x)$
can be written as $|a|\simeq \alpha t-\beta(v/J)$. At the
Rokhsar-Kivelson point $v/J=1$ and for $t=0$, we have that $e(x)=0$, which
implies $\beta(1)=0$ and the vanishing of $t^c$ for $h \to \Delta_B$. 
Away from the Rokhsar-Kivelson point, $\beta(v/J)$ is no longer vanishing and, 
from the fact that $\Delta_{B,c}^{\textrm{eff}}=2|a|$ (see above), we obtain
$t^c (h)\simeq (\Delta_B -h)/(2\alpha) + \beta/\alpha$,
that explains the behavior of the curves in Fig.~\ref{fig3}. 

Lastly, we note that the neglected terms 3) and 4) (as well as a direct 
exchange of two neighboring up and down spinons) may produce a finite density 
of minority down spinons which in turn might induce transverse spin-spin order.
However, we believe that transverse magnetic ordering would not change our 
qualitative picture. 

In summary, a simple extension of the QDM including spinon doping is 
introduced to account for the Zeeman effect of an applied magnetic field.
This model involves two semi-phenomenological parameters, the hopping 
amplitude $t$ for a spinon to hop (and exchange with a dimer) and the 
{\it bare} energy cost $\Delta_{B}$ to break up a dimer into two spinons.
Interestingly, the phase diagram of this model can be obtained from large-scale
Green's function Monte Carlo methods~\cite{ralko4} showing, as a function of 
the field, either a first order transition (square lattice) or a new type 
of Quantum Critical Point (triangular lattice) characterized by 
Bose-Condensation of spinons. 

F.B. acknowledges partial support from CNR-INFM and A.R. and D.P acknowledge
support from the Agence Nationale de la Recherche (France).


\begin{thebibliography}{99}
\bibitem{explain} Several microscopic models have been shown to
   be well described by the QDM e.g. the spin-orbital model describing LiNiO$_2$ compound 
   (see Ref.~\onlinecite{vernay}).
\bibitem{nota2spin} In the case with two spinons, there is no difference 
   between $S=0$ and $S=1$, since the spinons move on opposite
   sublattices.
\end{thebibliography}
\end{document}